# THE ORIGIN OF INTERGALACTIC THERMONUCLEAR SUPERNOVAE

A. G. Kuranov* and K. A. Postnov**

*Sternberg Astronomical Institute, Universitetskii pr. 13, Moscow, 119992 Russia*

**Abstract**—The population synthesis method is used to study the possibility of explaining the appreciable fraction of the intergalactic type-Ia supernovae (SN Ia), $20^{+12}_{-15}\%$, observed in galaxy clusters (Gal-Yam *et al.* 2003) when close white dwarf binaries merge in the cores of globular clusters. In a typical globular cluster, the number of merging double white dwarfs does not exceed $\sim 10^{-13}$ per year per average cluster star in the entire evolution time of the cluster, which is a factor of $\sim 3$ higher than that in a Milky-Way-type spiral galaxy. From 5 to 30% of the merging white dwarfs are dynamically expelled from the cluster with barycenter velocities up to 150 km s$^{-1}$. SN Ia explosions during the mergers of double white dwarfs in dense star clusters may account for $\sim 1\%$ of the total rate of thermonuclear supernovae in the central parts of galaxy clusters if the baryon mass fraction in such star clusters is $\sim 0.3\%$.

## INTRODUCTION

SN Ia (thermonuclear supernovae) are among the most important objects in modern astrophysics, because the maximum of their optical light curves has a small spread. Therefore, they can be used[1] as a standard candle in modern cosmology, and fundamental conclusions about the expansion kinematics of the Universe can be reached (Riess *et al.* 1998; Perlmutter *et al.* 1999). The physical cause of an SN Ia explosion is the thermonuclear explosion of a white dwarf (WD) with a mass close to the Chandrasekhar limit (Hoyle and Fauler 1960). This idea is fully confirmed by detailed numerical calculations of the evolution of a white dwarf in a binary system (see, e.g., Dunina-Barkovskata *et al.* 2001). The WD mass can increase during mass transfer between the components in a close binary, which determines the astrophysical sources of such supernovae—double WDs that merge through gravitational radiation (Iben and Tutukov 1984; Webbink 1984) or accreting WDs in semidetached or symbiotic binaries with a second nondegenerate component (Whelan and Iben 1973; Nomoto 1982). SN Ia are observed in galaxies of all morphological types (van den Berg *et al.* 2003), and their rate (in SNU units) depends weakly on the type of galaxy (see Capellaro *et al.* (1997) and more recent papers of Capellaro's group).

At present, it is not completely clear which of these scenarios (or both) make the largest contribution to the SN Ia rate, and a considerable observational effort is being made to find the Galactic population of double WDs with a total mass close to or larger than the Chandrasekhar limit (e.g., the SPY project by Napiwotzki *et al.* 2003). We will adhere to the scenario of merging double WDs as the progenitors of type-Ia supernovae. The recent discovery of hydrogen emission lines in the spectrum of SN 2002ic (Hamyu *et al.* 2003) cannot rule out this formation channel of thermonuclear supernovae, because such SN Ia are atypical; their origin is the subject of debate (e.g., Livio and Riess 2003; Chugai and Yungelson 2003).

A purposeful search for supernovae in Abell clusters of galaxies (the WOOTS project; Gal-Yam *et al.* 2003) has revealed that two of the seven SN Ia found (SN 1998fc in Abell 403 at $z = 0.1$ and SN 2001al in Abell 2122/4 at $z = 0.066$)) have no obvious host galaxies and can be associated only with faint dwarf galaxies with $M_R > -12^m$. In reality, in both cases, the supernovae are projected onto the halos of the central cD galaxies in the corresponding clusters. However, the authors rejected the possibility that they belong to these galaxies because of the large radial-velocity difference (750−2000 km s$^{-1}$) between the supernovae and the corresponding cD galaxies.

Assuming a modified Schechter luminosity function for the galaxies in the Virgo cluster (Trentham and Tully 2002), the luminosity fraction of the dwarf galaxies in the clusters under study should have been less than 0.3%. If the supernova rate is proportional to luminosity, and if the fraction of the supernovae without obvious host galaxies is on the order of 10%, this leads to their $\sim 30$-fold excess among such dwarf galaxies compared to high-mass galaxies. Therefore, these supernovae are believed to have exploded outside galaxies (intergalactic supernovae) as a result of the evolution of the intergalactic stellar population. It is concluded from these observations that the fraction of the intergalactic SN Ia is $20^{+12}_{-15}\%$, in accordance with the theoretically expected fraction of the stars outside galaxies due to tidal interaction between cluster galaxies. The authors themselves (Gal-Yam et al. 2003) point out a low statistical significance of the result obtained, and such a high percentage of the extragalactic SN Ia should be confirmed by further observations.

---

*E-mail: alex@xray.sai.msu.ru
**E-mail: pk@sai.msu.ru

[1] However, with certain reservations, up to 40% of SN Ia may be peculiar (Lee *et al.* 2001), plus a significant spread of theoretical light curves (Sorokina *et al.* 2000).

There is a potential channel for the formation of intergalactic supernovae through the dynamical expulsion of close WD pairs from the system of galactic globular clusters (GCs). The stellar mass fraction in the observed GCs is $\sim 0.1\%$ of the visible mass of the galaxies. However, dynamical interactions between stars in dense cluster cores increase the merger rate of close pairs with white dwarfs. Some of them can be expelled from the clusters during tripple collisions, and the assumption that the rate of thermonuclear supernovae is proportional to luminosity for dense clusters becomes invalid. The numerical calculations of the WD evolution in open star clusters by Shara and Hurley (2002) confirmed the significant increase in the formation rate of close WD pairs through dynamical interactions.

The goal of our study is to quantitatively estimate the merger rate of double WDs in the dense cores of star clusters by using the method of population synthesis of the evolution of binary stars with a semi-analytical allowance made for their dynamical interactions with single stars in the cluster core. We show that the the merger rate of double WDs in GCs per average cluster star is a factor of $\sim 3$ higher than the merger rate of double WDs in spiral galaxies with continuous star formation. If $\sim 0.3\%$ of the baryon mass in the cluster centers is concentrated in virialized dense star clusters with masses of $10^5-10^8 M_\odot$, then $\sim 1\%$ of SN Ia in the cluster centers can be naturally explained without additionally assuming the presence of 10% of the intergalactic stellar population.

## THE MODEL

*The Structure and Evolution of a Globular Cluster*

In our calculations, we used a Michie–King model (Michie 1963) to describe the structure of the globular cluster. The stellar population was broken down into subsystems (depending on their mass); their space density $\rho_\alpha$ was fitted by a power law:

$$\rho_\alpha(r) = \begin{cases} \rho_{c_\alpha}, & r \leq r_{c_\alpha}, \\ \rho_{c_\alpha}(r/r_{c_\alpha})^{-2}, & r_{c_\alpha} < r \leq r_{h_\alpha}, \\ \rho_{h_\alpha}(r/r_{h_\alpha})^{-4}, & r_{h_\alpha} < r \leq r_t, \end{cases} \quad (1)$$

where $r_{c_\alpha}$ and $r_{h_\alpha}$ are, respectively, the radii of the core and the sphere within which half the mass of the $\alpha$ subsystem of stars is contained; and $r_t$ is the tidal radius of the cluster. Thus, $\rho_{h_\alpha} = \rho_{c_\alpha}(r_{h_\alpha}/r_{c_\alpha})^{-2}$ and $\rho_{c_\alpha} = M_{\text{tot}_\alpha}/\left[8\pi r_{c_\alpha}^2 \left(r_{h_\alpha} - \frac{2}{3}r_{c_\alpha}\right)\right]$. Under the energy equipartition condition for the GC stars, we may assume that

$$r_{c_\alpha} = \sqrt{\frac{\bar{m}_c}{m_\alpha}} r_c,$$

where $\bar{m}_c$ is the mean mass of the stars in the GC core, $r_c = \sqrt{\frac{3 v_m(0)^2}{4\pi G \rho_c}}$ is the radius of the GC core, $v_m$ is the rms space velocity, and $\rho_c$ is the central number density of the stars:

$$\rho_c = \sum_\alpha \rho_{c_\alpha}. \quad (2)$$

The stellar velocities can be described by the truncated Maxwellian distribution

$$g(E) = \begin{cases} Ke^{-\gamma J^2}[e^{-\beta E} - e^{-\beta E_t}], & E < E_t, \quad J < J_c(E), \\ 0, & E > E_t, \end{cases} \quad (3)$$

where $E$ and $J$ are the energy and the angular momentum (per unit mass), respectively; and $J_c(E)$ is the value of $J$ for a star in a circular orbit with energy $E$. Stars with $E > E_t$ are assumed to have escaped from the cluster.

Detailed calculations of the evolution of the standard model reveal a collapse of the inner core and an expansion of the outer layers. As follows from numerical calculations (see, e.g., Kim *et al.* 1998), the core contraction at late stages asymptotically approaches a power law:

$$\rho_c(t) \sim \rho_c(0)(1 - t/t_{\text{coll}})^{-1.2}, v_m(t) \quad (4)$$
$$\sim v_m(0)(1 - t/t_{\text{coll}})^{-0.12},$$

where $t$ is the current time, and $t_{\text{coll}}$ is the collapse time of the cluster.

As the GC core contracts, the central number density of the stars rapidly increases, which causes the formation rate and the binding energy of binary stars to increase. Once the initial collapse has stopped, one might expect an expansion of the cluster core due to the tidal formation of binary stars. For some time, the newly forming binaries will be the energy source that maintains the expansion of the inner cluster regions. The expansion can be described by the following asymptotic formulas (see Kim *et al.* 1998):

$$\rho_c(t) \sim t^{-2}, v_m(t) \sim t^{-0.32}. \quad (5)$$

We used several GC models that differed in central density $\rho_c$ and radius $r_c$ at the time of the collapse. In all models, the collapse time was taken to be $t_{\text{coll}} = 7 \times 10^9$ yr, the central potential was $W_0 = -\beta\phi(0) = 9$ ($\beta \equiv 3/v_m^2$, $\phi(r)$ is the gravitational potential), the total number of stars was $10^6$, and the fraction of the initially binary stars was assumed to be 30%.

The last two parameters specify the normalization of our calculations. For the adopted Salpeter initial mass function with a minimum mass of the forming stars of $0.1 M_\odot$, the mean stellar mass in the GC is $0.38 M_\odot$, and the total mass of the GC of $10^6$ stars is $\sim 4 \times 10^5 M_\odot$. Thus, the results presented below pertain to modeling the evolution of $3 \times 10^5$ binaries in a GC with a mass of $1.3 \times 10^5 M_\odot$.

Here, it is pertinent to make a note on the chosen 30% fraction of the initially binary stars in the GC. It follows from Hubble Space Telescope observations of GCs (Rubinstein and Bailyn 1997) that the fraction of the main-sequence binary stars in the GC with a collapsed core NGC 6752 may range from 15 to 38%. In NGC 288, this fraction is estimated to be in the range from 8 to 38% (Bellazzini *et al.* 2002). A numerical analysis of the evolution of binaries in a GC (Fregeau *et al.* 2003) indicates that the fraction of the binaries with respect to their initial number after the core collapse ranges (depending on the cluster model) from $\sim 0$ to 35% (the table in the paper by Fregeau *et al.* 2003). This is attributable to the effective dynamical disruption of wide pairs and to the mergers of close systems. Note also that the above authors assumed up to 30% of the initially hard binaries in the GC. We choose an upper limit of $1000 R_\odot$ for the initial binary semiaxes, which corresponds to the criteria by Fregeau *et al.* (2003). Thus, the high (at first glance) percentage of the initially binary systems that we adopted is not overestimated and is consistent with available observations. Unfortunately, we cannot accurately estimate the number of newly forming binaries due to tidal captures. In any case, their percentage is much lower than the fraction of the initially binary systems.

*The Evolution of a Binary in a Globular Cluster*

The Scenario Machine population synthesis code (Lipunov *et al.* 1996) was used to model the evolution of binary and single stars. We used the initial distributions

$$f(\lg a) = \text{const}, \quad (6)$$

$$\max\left\{\begin{array}{c} 10 R_\odot \\ R_L(M_1) \end{array}\right\} < a < 10^3 R_\odot,$$

for the semimajor axes of the binaries and

$$f(M_1) \propto M_1^{-2.35}, \quad 0.1 M_\odot < M_1 < 120 M_\odot \quad (7)$$

for the mass of the primary component. Here, $R_L(M_1)$ is the Roche lobe radius for the primary component. For the component mass ratio, we used a power-law distribution:

$$f(q) \propto q^{\alpha_q}, \quad q = M_2/M_1 < 1. \quad (8)$$

Here, $\alpha_q$ is the parameter of the distribution in component mass ratio. In our calculations, we varied it between 0 and 1.

Another important parameter of the evolution of binaries that significantly affects the results of all such calculations is the energy transfer efficiency at the common-envelope stage, $\alpha_{\text{CE}}$. Having been determined both by Lipunov *et al.* (1996) and in other studies of this group, it was assumed to be $\alpha_{\text{CE}} = 0.5$. This value corresponds to the (standard) value of $\alpha_{\text{CE}} = 1$ in the determination used by Hurley *et al.* (2002), in which the common envelope was ejected only through the transfer of the gravitational energy of orbital motion of the approaching components to it.

When describing the dynamical interactions between cluster stars, we took into account the following processes in the code:

(1) The passages that led to changes in the orbital parameters of binaries (semimajor axis and eccentricity).

(2) The passages that led to star exchanges.

(3) The passages that led to the disruption of binaries.

(4) The interactions between single stars that led to the formation of binaries.

The corresponding cross sections $\sigma_{ij}$ for these processes were numerically calculated by several authors (see Heggei *et al.* 1996; Mikkola 1984; Kim and Lee 1999).

The rate of interaction between an individual binary $b$ at distance $r$ from the GC center with an $\alpha$ subsystem of stars with a space density $n_\alpha(r)$ at time $t$ is given by the formula

$$\mathcal{R}_{\alpha b}(t) = n_\alpha(r,t) v_m(r,t) \sigma_{\alpha b}.$$

The determination of the binary position in the GC and $\mathcal{R}$ is described in more detail in the Appendix.

To draw a change in the binding energy $\Delta E_b$ and orbital eccentricity $e$ of a binary during the close passage of a third star, we used the following formulas for the differential interaction cross section (Heggei *et al.* 1993; Davis *et al.* 1992):

$$\frac{d\sigma}{dX} = k\pi a_0^2 \frac{X^{-\Delta}(1+X)^{-4.5+\Delta}}{V^2}, \quad (9)$$

$$\frac{d\sigma_e}{de} \sim e, \quad (10)$$

where $X = \Delta E_b / E_b$, and the parameter $\Delta$ depends on the relative velocity of the approaching stars. The mean value is $\langle X \rangle = 0.4$.

The change in binding energy during a tripple collision results in momentum transfer to the binary barycenter. Thus, it determines the possibility of the binary being expelled from the cluster through such a collision and the distance to which the escaped binary can go before its merger.

### RESULTS OF THE CALCULATIONS

Since the expulsion of merging double WDs from a cluster is directly related to the dynamical interactions between stars, one might expect the number of such systems $N_{\rm WD^2}^{ej}$ in the evolution time to be determined by $\Gamma$:

$$\Gamma \propto \int \mathcal{R}_{\rm WD^2} n_{\rm WD^2} dV \sim \frac{\rho_0^2 r_c^3}{v_m} \sim \rho_0^{1.5} r_c^2,$$

where the integration is over the entire GC volume.

The calculated numbers of merging WD pairs for various GC models are given in the table. We see from this table that the choice of a GC model (parameter $\Gamma$) most strongly affects precisely the number of WDs dynamically expelled from the cluster: $N_{\rm WD^2}^{ej} \propto \Gamma^{0.65}$. A similar power law also follows from observations for X-ray sources in the GC. In this case, the power-law index for X-ray sources, which are associated with higher-mass binaries, is close to unity: $N_{\rm X-ray} \propto \Gamma^{0.74\pm0.36}$ (Pooley *et al.* 2003).

The merger rate of double WDs (SN Ia candidates) is plotted against time for one model (C) in

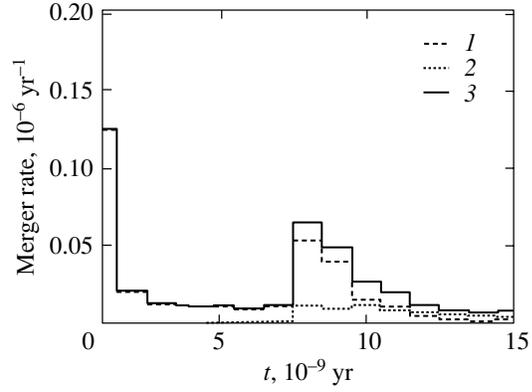

**Fig. 1.** Merger rate of double WDs in a GC (model C, $\alpha_q = 1$) versus time, in units of $10^{-6}$ yr$^{-1}$. The increase in merger rate at $t \sim 7 \times 10^9$ yr corresponds to an increase in $\Gamma$ during the GC core collapse: *1*—for systems merging within the GC; *2*—for systems expelled from the GC; and *3*—the total merger rate of double WDs in the GC.

The number of merging double white dwarfs in a GC of $10^6$ stars with a total mass of $4 \times 10^5 M_\odot$ in an evolution time of $1.5 \times 10^{10}$ yr

| Model | $\rho_c$, pc$^{-3}$ | $r_c$, pc | $\Gamma$ | $N_{\rm WD^2}$ | $N_{\rm WD^2}^{\rm GC}$ | $N_{\rm WD^2}^{\rm ej}$ | $\nu$ |
|---|---|---|---|---|---|---|---|
| A | $10^5$ | 0.1 | 1 | 108 | 98 | 10 | 0.09 |
|   |        |     |   | 228 | 216 | 12 | 0.05 |
| B | $10^5$ | 0.3 | 10 | 120 | 108 | 12 | 0.10 |
|   |        |     |    | 252 | 216 | 36 | 0.14 |
| C | $10^6$ | 0.1 | 32 | 138 | 114 | 24 | 0.17 |
|   |        |     |    | 330 | 264 | 66 | 0.20 |
| D | $10^7$ | 0.03 | 90 | 174 | 132 | 42 | 0.24 |
|   |        |      |    | 381 | 261 | 120 | 0.31 |

Note. The initial number of binaries is $3 \times 10^5$. The GC parameters are given at the time of the core collapse ($7 \times 10^9$ yr). The value of $\Gamma$ was normalized in such a way that $\Gamma = 1$ for model A. $N_{\rm WD^2}$ is the total number of merging double WDs in the entire evolution time; $N_{\rm WD^2}^{\rm GC}$ and $N_{\rm WD^2}^{\rm ej}$ are the numbers of merging double WDs within the GC and expelled from the GC, respectively; and $\nu$ is the fraction of the expelled merging double WDs. For each model, the results of our calculations are presented for two values of $\alpha_q$: $\alpha_q = 0$ (upper rows) and $\alpha_q = 1$ (lower rows) (see formula (8)).

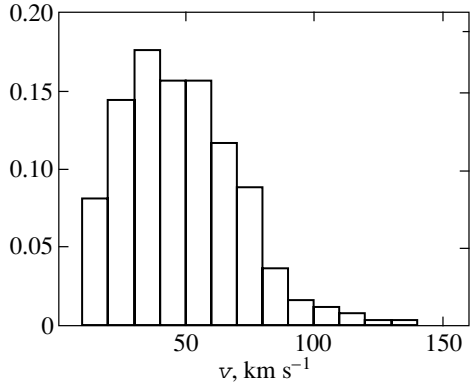

**Fig. 2.** Histogram (normalized to unity) illustrating the distribution of WD pairs expelled from the cluster in their barycenter velocities (model C).

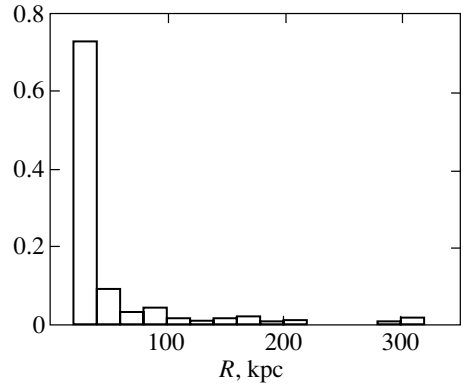

**Fig. 3.** Histogram (normalized to unity) illustrating the distribution of WD pairs expelled from the cluster in distance from the center of an isolated globular cluster at the time of their merger (model C).

Fig. 1. The increase in merger rate at time $t \sim 7 \times 10^9$ yr corresponds to a sharp increase in $\Gamma$ during the evolution of the cluster (GC core collapse).

The ratio of the total number of mergers to the number of evolving binaries may serve as a characteristic of the efficiency of the merger rate of double WDs that is independent of the adopted normalization (the total number of stars in the cluster and the fraction of the initially binary stars). Taking 300 mergers per $3 \times 10^5$ binaries (table) as a typical value, we obtain an efficiency of $10^{-3}$ per initial binary system. This value is half the value obtained by Shara and Hurley (2002) (2002) in their numerical simulations and can be explained, in particular, by the higher efficiency of the common envelope ($\alpha_{CE} = 3$) in these numerical simulations.

Note also that the total number of merging WDs as a result of the GC evolution, $N_{WD^2}$, changes by a factor of only 1.5 for different GC models and is 100–500 per cluster; most (70%) of the mergers occur after the collapse of the cluster core. The mean merger rate of double WDs in the past 5 Gyr is then $\sim 10^{-13}$ per year per average cluster star, which is a factor of $\sim 3$ higher than the estimated merger rate of double WDs in our Galaxy (Nelemans et al. 2001; Hurley et al. 2002).

The distributions of WD pairs expelled from the cluster in their barycenter velocities and distances from the center of the host GC are shown in Figs. 2 and 3. As we see from Fig. 2, the fraction of the stars with escape velocities of about 100 km s$^{-1}$ is small (a few percent).

## DISCUSSION
*The Escape of Merging Double WDs from the System of Galactic Globular Clusters*

**Our Galaxy.** The GC system of our Galaxy extends to distances as large as 100 kpc above the Galactic plane and has a velocity dispersion of about 150 km s$^{-1}$ (see, e.g., the catalogs by Kukarkin (1974) and Harris (1996)). The escape velocity from the Galactic potential is $\sim 550$ km s$^{-1}$ in the Galactic plane and decreases to $\sim 100$ km s$^{-1}$ at distances of $\sim 100$ kpc from the Galactic center (accurate estimates are difficult to obtain, because the form of the Galactic potential is not known at such distances). The escape of the stars expelled from the GC is preferred at the apocenter of the cluster orbit, near which the cluster is located for the longest time. Clearly, not all of the binaries expelled from the cluster will be able to escape from the host galaxy, but will remain in its halo. Thus, the total number of WD pairs escaping from the cluster gives an *upper* limit on the number of intergalactic type-Ia supernovae. The rough upper limit on the probable number of intergalactic SN Ia, $N_{\rm SN\ Ia}$, that emerge during the mergers of double WDs expelled from the GC in a cluster of $N_{\rm gal}$ galaxies is

$$N_{\rm SN\ Ia} = N_{\rm WD^2} \times N_{\rm GC} \times N_{\rm gal}$$
$$\sim 100 \times 500 \times 10^3 = 5 \times 10^7$$

($N_{\rm GC}$ is the number of GCs in an average galaxy). In a characteristic time of $10^{10}$ yr, the formation rate of SNe Ia through this channel throughout the cluster is $\sim 5 \times 10^{-3}$ per year, which is comparable to the merger rate of double white dwarfs with a total mass exceeding the Chandrasekhar limit in a *single* galaxy (Nelemans et al. 2001). Even this upper limit is an order of magnitude smaller than the high fraction of the extragalactic SNe Ia deduced from observations.

**Dwarf elliptical galaxies.** We may also assume that dwarf elliptical dE galaxies (the fraction of which in a cluster can be significant) with a mass up to $M \sim 10^8 M_\odot$ are the main source of observed intergalactic supernovae. The central number densities of the stars in such galaxies are comparable to those in dense GC cores. The escape velocity from a cluster at a constant star space density is $v_e \propto M^{1/3}$. Therefore, applying our calculations to a system of $10^8$ stars, we find that the number of escaping WD pairs from this system is $10^8/10^6 \times f(> v_e) \times 100 \sim 100$ (in this estimate, the fraction of the binaries with barycenter velocities higher than the escape velocity from the cluster, $v_e \sim 100\,{\rm km\,s^{-1}}$, was taken to be $f(> v_e) \sim 1/100$), which is also insufficient to explain the truly intergalactic supernovae.

**cD galaxies.** Let us now turn to the central cD galaxy as the source of intergalactic supernovae. Although SN 1998fc in Abell 403 and SN 2001al in Abell 2122/4 are projected onto the halos of the central cD galaxies (the projected distance is 160 kpc), we rule out the possibility that they belong to these galaxies because of the radial-velocity difference (Gal-Yam et al. 2003). The giant elliptical galaxies in the cluster centers are known to have a huge system of globular clusters (up to 10000). Using the results obtained in our calculations ($\sim 300$ mergers of double WDs in cluster cores and $\sim 100$ merging pairs escaped after the core collapse), we obtain an average estimate of the maximum SN Ia rate due to the WD evolution in the GC in projection onto the halo of the cD galaxy, $400 \times 10^4/(5 \times 10^9) \sim 10^{-3}$ per year. Only a few percent of such double WDs can have barycenter velocities high enough to escape from the galaxy; they cannot explain the observations.

*The Mergers of Double WDs in Virialized Star Clusters at the Centers of Galaxy Clusters*

Is there an alternative explanation for the observations by Gal-Yam et al. (2003) that does not appeal to the evolution of $10-20\%$ of the stars outside galaxies? In our view, the most plausible explanation for intergalactic supernovae is the merger of WD pairs in a virialized system of star clusters in the central regions of galaxy clusters. It follows from observations of the central parts of the Virgo cluster and from simulations (see the recent review article by Lee (2002) and references therein) that a huge system of blue GCs with a large velocity dispersion must be formed during the formation of giant elliptical galaxies. Recent observations of intergalactic (up to 250 kpc from the center) GCs in the nearby Hydra I and Centaurus clusters (Hilker 2002) lead to a similar conclusion. An enhanced density of star clusters in the central regions of galaxy clusters naturally arises during the merger of the galaxies that form giant cd galaxies as well as through the tidal capture of the GC system of the galaxies that pass near the cluster center. Assuming that 0.3% of the luminosity (and the baryon mass) in the cluster centers is concentrated in dense collapsed stellar structures and taking into account the threefold increase in the merger rate of double WDs in such clusters, we naturally obtain $\sim 1\%$ of the observed SN Ia rate *outside* the visible host galaxies. Thus, this model predicts the existence of faint ($\langle M_R \rangle \sim -10^m$) *host clusters* for SN Ia. It is hoped that the purposeful search for type-Ia supernovae being conducted at present as part of various projects (see, e.g., Tonry et al. 2003) will improve the existing statistics and will reveal the faint host galaxies and astrophysical sources of intergalactic SN Ia. The absence of host clusters with $M_R > -8^m \ldots -10^m$ may serve as a critical argument for choosing the model of intergalactic type-Ia supernovae from intergalactic stars.

CONCLUSIONS

We have considered the possible formation channel of intergalactic SN Ia during the mergers of double white dwarfs that evolve and are expelled from dense star clusters (GCs, the nuclei of dwarf elliptical galaxies, etc.) when they dynamically interact with a third star (tripple collisions). The population synthesis method was used to model the formation and evolution of close double white dwarfs in cluster cores by taking into account the time evolution of their physical parameters and to calculate the rates of escape of such objects from clusters. We obtained their distributions in escape velocities from clusters, in merger rates, and in distances from the host cluster after escape. We showed that the mean merger rate of double WDs after the core collapse could reach $\sim 10^{-13}$ per year per average GC star. In a typical GC, the number of expelled WD pairs after the GC core collapse is $\sim 100$. For a galaxy cluster with several thousand members, this estimate yields a formation rate of intergalactic thermonuclear supernovae $< 0.005$ per year, which is much lower than the observed high rate of intergalactic supernovae of this type. We hypothesize that the possible observed high percentage of intergalactic SN Ia without any obvious host galaxies can be explained in part by the evolution of double WDs in dense star clusters. A system of such clusters naturally arises near the center of galaxy clusters during the formation of giant elliptical galaxies and, subsequently, during the tidal interaction between the galaxies that pass near the cluster center. This hypothesis can be verified through observations of the faint host clusters of SN Ia in nearby galaxy clusters.

ACKNOWLEDGMENTS

We wish to thank A.S. Rastorguev and O.K. Sil'chenko for helpful discussions and L.R Yungelson for valuable remarks. This work was supported by the Russian Foundation for Basic Research (project nos. 03-02-16110a and 03-02-06733mas).

*APPENDIX*

The position of a binary in a globular cluster.

The relative motions of stars in a GC produce continuous fluctuations of the gravitational field. In turn, these fluctuations cause the magnitude and direction of the velocity of each star to change. As a result, the energy $E = v^2/2 + \phi(r)$ and the angular momentum $J = r \times v_t$ of the star will change. When calculating each evolutionary track, the changes in these parameters in a time interval $\Delta t$ due to the gravitational interaction with the surrounding stars can be calculated by using the formula

$$\Delta E = n\epsilon_1 + n^{1/2} y_1 \epsilon_2^2,$$
$$\Delta J = n j_1 + n^{1/2} y_2 j_2^2,$$
$$\Delta t = n P(E, J),$$

where $y_1$ and $y_2$, the random numbers drawn from a normal distribution with a mean $\langle y_1 \rangle = \langle y_2 \rangle = 0$ and a variance $\langle y_1^2 \rangle = \langle y_2^2 \rangle = 1$, are chosen from the correlation condition in such a way that

$$\langle y_1 y_2 \rangle = \xi^2 / \epsilon_2 j_2.$$

We used a method with averaging over the orbit similar to the method by Shapiro and Marchant (1978) and Marchant and Shapiro (1979) to determine the coefficients $\epsilon_1$ and $\epsilon_{2,1}$:

$$\epsilon_1 = 2 \int_{r_p}^{r_a} \langle \Delta E \rangle dr / v_r,$$
$$\epsilon_2^2 = 2 \int_{r_p}^{r_a} \langle \Delta E^2 \rangle dr / v_r,$$
$$j_1 = 2 \int_{r_p}^{r_a} \langle \Delta J \rangle dr / v_r,$$
$$j_2^2 = 2 \int_{r_p}^{r_a} \langle \Delta J^2 \rangle dr / v_r,$$
$$\xi^2 = 2 \int_{r_p}^{r_a} \langle \Delta E \Delta J \rangle dr / v_r.$$

Here, $r_p$ and $r_a$ are the pericenter and apocenter of the stellar orbit in the cluster. These coefficients are the mean rates of change in $E$ and $J$ due to several approaches over the orbital period and the mean cumulative values over the orbital period.

The diffusion coefficients that contain $E$ and $J$ can be expressed in terms of $\langle \Delta v_\| \rangle, \langle \Delta v_\|^2 \rangle$ and $\langle (\Delta v_\perp)^2 \rangle$ are the locally calculated velocity diffusion coefficients:

$$\langle \Delta E \rangle = v \langle \Delta v_\| \rangle + \frac{1}{2} \langle \Delta v_\perp^2 \rangle + \frac{1}{2} \langle \Delta v_\|^2 \rangle,$$
$$\langle \Delta E^2 \rangle = v^2 \langle \Delta v_\|^2 \rangle,$$
$$\langle \Delta J \rangle = \frac{J}{v} \langle \Delta v_\| \rangle + \frac{r^2}{4J} \langle \Delta v_\perp^2 \rangle,$$
$$\langle \Delta J^2 \rangle = \frac{J^2}{v^2} \langle \Delta v_\|^2 \rangle + \frac{1}{2}(r^2 - \frac{J^2}{v^2}) \langle \Delta v_\perp^2 \rangle,$$
$$\langle \Delta E \Delta J \rangle = J \langle \Delta v_\|^2 \rangle$$

For a Maxwellian velocity distribution of the field stars, the diffusion coefficients reduce to standard form (Spitzer 1987):

$$\langle \Delta v_\| \rangle = -2 \left(1 + \frac{m}{m_f}\right) n_f \Gamma k^2 G(x),$$
$$\langle (\Delta v_\|)^2 \rangle = 2 n_f \Gamma k \frac{G(x)}{x},$$
$$\langle (\Delta v_\perp)^2 \rangle = 2 n_f \Gamma k \frac{\Phi(x) - G(x)}{x},$$

where $\Phi(x)$ is the error function

$$\Phi(x) = \frac{2}{\pi^{1/2}} \int_0^x \exp -y^2 dy$$

and

$$G(x) \equiv \frac{\Phi(x) - x\Phi'(x)}{2x^2}.$$

Using $E$ and $J$, we can determine the orbital parameters of the star within the GC: the orbital period

$$P(E, J) = 2 \int_{r_p}^{r_a} dr / v_r,$$

the pericenter $r_p$ and the apocenter $r_a$ of the stellar orbit in the cluster from the condition

$$v_r^2 = 2E - 2\phi(r) - J^2/r^2 = 0.$$

The rate of interaction between a binary $b$ and an $\alpha$ subsystem of stars at time $t$ is given by the formula

$$\mathcal{R}_{\alpha b}(t) = \sigma_{\alpha b} \times \frac{2}{P(E, J)} \int_{r_p}^{r_a} n_\alpha(r, t) v(r, t) / v_r dr.$$